\documentclass[letterpaper, 10 pt, conference]{ieeeconf}  % Comment this line out if you need a4paper

\IEEEoverridecommandlockouts                              % This command is only needed if 
\usepackage{blindtext}

\usepackage{verbatim,multirow,hhline}
\usepackage[usenames,dvipsnames]{xcolor}

  \usepackage[pdftex]{graphicx}
\usepackage[cmex10]{amsmath}
\usepackage{subfigure}
\usepackage{tikz}
\usepackage[utf8]{inputenc}
\usepackage{amsfonts}
\usepackage{amssymb}
\usepackage{multirow}
\usetikzlibrary{shapes,arrows, fit}
\usetikzlibrary{positioning}

\tikzstyle{response} = [cloud, circle, draw, fill=blue!10, 
    text width=5em, text centered,  minimum height=3em]

\tikzstyle{level1} = [cloud, circle, draw, fill=white!10, 
    text width=4em, text centered,  minimum height=2em] 
    
\tikzstyle{desire} = [cloud, circle, draw, fill=red!10, 
    text width=4em, text centered,  minimum height=2em] 

\tikzstyle{level2} = [cloud, circle, draw, fill=red!10, 
    text width=4em, text centered,  minimum height=2em]

\tikzstyle{arrow} = [ultra thick,->,>=stealth]
\tikzstyle{a} = [rectangle, draw, minimum height=17em, minimum width=15em]
\title{\LARGE \bf
Context-Aware Recursive Bayesian Graph Traversal in BCIs*
}

\author{Seyed Sadegh Mohseni Salehi$^{2}$, Mohammad~Moghadamfalahi, Hooman Nezamfar,\\ Marzieh Haghighi and Deniz~Erdogmus$^{1}$% <-this % stops a space
\thanks{*Our work is supported by NSF (IIS-1149570, CNS-1544895), NIDLRR (90RE5017-02-01), and NIH (R01DC009834). Relevant code and data will be disseminated via Northeastern University Digital Repository Service, in collection NEU/COE/ECE/CSL (permalink: http://hdl.handle.net/2047/D20199232).}% <-this % stops a space
\thanks{$^{1}$ {Electrical and Computer Engineering Department, Northeastern University, Boston}}%
\thanks{$^{2}$ {\tt\small ssalehi@ece.neu.edu}}
}

\begin{document}

\maketitle
\thispagestyle{empty}
\pagestyle{empty}

%%%%%%%%%%%%%%%%%%%%%%%%%%%%%%%%%%%%%%%%%%%%%%%%%%%%%%%%%%%%%%%%%%%%%%%%%%%%%%%%
\begin{abstract}  
Noninvasive brain computer interfaces~(BCI), and more specifically Electroencephalography~(EEG) based systems for intent detection need to compensate for the low signal to noise ratio of EEG signals. In many applications, the temporal dependency information from consecutive decisions and contextual data can be used to provide a prior probability for the upcoming decision. In this study we proposed two probabilistic graphical models~(PGMs), using context information and previously observed EEG evidences to estimate a probability distribution over the decision space in graph based decision-making mechanism. In this approach, user moves a pointer to the desired vertex in the graph in which each vertex represents an action. To select a vertex, a “Select” command, or a proposed probabilistic \emph{Selection} criterion~(PSC) can be used to automatically detect the user intended vertex. Performance of different PGMs and \emph{Selection} criteria combinations are compared over a keyboard based on a graph layout. Based on the simulation results, probabilistic \emph{Selection} criterion along with the probabilistic graphical model provides the highest performance boost for individuals with pour calibration performance and achieving the same performance for individuals with high calibration performance.

\end{abstract}
%%%%%%%%%%%%%%%%%%%%%%%%%%%%%%%%%%%%%%%%%%%%%%%%%%%%%%%%%%%%%%%%%%%%%%%%%%%%%%%%
\section{Introduction}
 BCIs promise to provide a new solution for people with major speech and muscle impairments to interact with their environment and communicate with others. EEG is a non-invasive method for recording brains electrical activity~\cite{curran2003learning}. Being non-invasive, EEG-based BCIs have attracted significant attention, to be become a viable and safe solution. 

P300 and Steady State Visually Evoked Potential (SSVEP) are more popular in BCI systems where the brevity of response time and the availability of several options are concerned. Researchers in the field have exploited these features to facilitate wheelchair navigation and develop spellers. In control and navigation applications SSVEP-based BCI systems are widely used due to their fast response. For instance, Muller et. al. proposed a four-command, SSVEP-based robotic wheelchair~\cite{muller2011using, nezamfar2013brain}. In another study, P300 and SSVEP were used in a hybrid system to control a wheelchair~\cite{li2013hybrid}. However, EEG is known to have low signal to noise ratio. Studies have shown that fusion of other information sources (e.g. contextual information and/or history of user's earlier selection) can enhance system performance~\cite{moghadamfalahi2015active, moghadamfalahi2015language}. % are limited in the sense that they only use brain signal to decide new commands but stop short of utilizing other existent sources of information.

In applications with more than two command, usage of the P300 component might not provide the best performance. Typically in these BCIs not only the system requires multiple trials of each target in order to make a confident decision; but also, the system needs to map many options onto two classes which can lead to significant delay in decision making. Many researchers employ a different approach in which, the user navigates a cursor in a gridded space of options to the desired node and make a selection~\cite{valbuena2008spelling, volosyak2011ssvep, chen2016music}. The Bremen BCI keyboard is an example of such systems in which, an optimized keyboard layout is presented to the user and they can perform letter by letter typing through navigating a cursor on a grid of characters~\cite{valbuena2008spelling, volosyak2011ssvep}. Similarly, Iwan et.al.~\cite{iwane2016spatial} used Error-related potential (ErrP) as a biological signal to navigate the cursor through the gridded space of options to select the desired option.
But most of these researches such as ~\cite{valbuena2008spelling,volosyak2011ssvep, iwane2016spatial}, do not use neither the layout information or the confidence by which earlier selections are performed. 

Our earlier work, Flashtype~\cite{nezamfar2016flashtype}, is utilizing a probabilistic graphical model which uses context information and the evidence accumulated from earlier decisions to calculate prior probability distribution over the state space. In this study, we propose a new probabilistic graphical model maintaining the use of context information and the evidence accumulated from earlier decisions. The introduced statistical inference mechanism provides a new \emph{Selection} criterion to estimate user intent. 
We report the performance of these models using our earlier designed BCI, Flashtype\texttrademark~\cite{nezamfar2016flashtype}. A simulation study using real code modulated VEP (c-VEP) data collected during a \emph{Calibration} session from seven subjects is used to compare possible performance improvements using the proposed methods.

\section{Methods}
\subsection{BCI Graph Navigation Framework}
\begin{figure}[h!]
        \begin{minipage}[b]{1.\linewidth}
        \centering
        \centerline{\includegraphics[width=6cm]{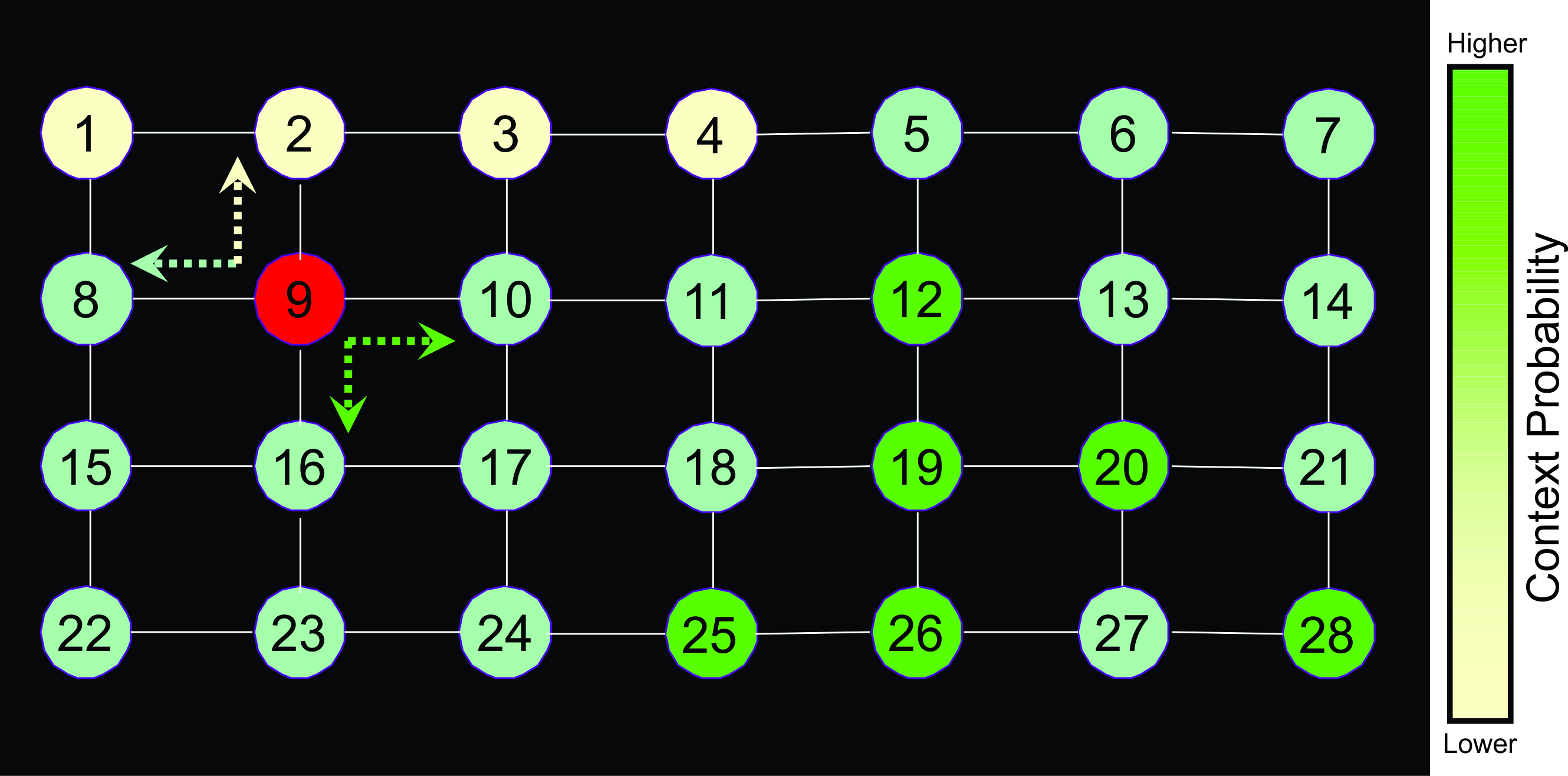}}
        \vspace{-15pt}
        \centerline{}\medskip
        \end{minipage}
        \vspace{-.8cm}
        \caption{{\footnotesize BCI navigation framework schematic.}}
        \label{fig:graph}
    \end{figure}

In this section a general framework for a BCI navigation system is described. As shown in Fig.~\ref{fig:graph}, a navigation graph contains a set of connected nodes. At each node an immediate action of moving to a neighbor node (from $m$ possible neighbor nodes in a connected graph with $n$ nodes) or current node selection can be taken. Therefore, a set of immediate actions are needed to converge to the target node through a specific path.
In a BCI navigation system, each immediate action of moving to a neighbor node or current node selection is taken by using subject's brain evidence and contextual information. In our system, each navigation sequence which is called an epoch contains a set of immediate actions through $t$ trials ($1\leq t \leq N_t$) for converging to the target node. At the end of each epoch, navigation sequence to the target node is finalized by a selection, based on a \emph{Selection} criterion. In this probabilistic framework, in each trial next neighbor node is decided by estimation of subjects intended action. In our system, this probabilistic estimation in epoch $e$ and trial $t$ is attained by fusion of (1) EEG evidence, (2) prior probabilities of each node being target and (3) history of previous actions and their probabilities in that epoch. 
\subsection{Graphical Models for Statistical Inference}
\label{sec:pgms}
%In a hierarchical/pointer-based decision-making mechanism the user navigates the pointer in the connected graph with n vertices (number of actions) of degree m (number of EEG classes or the cardinality of decision space) to choose from several actions. Each navigation sequence to a desired vertex is finalized by a selection, based on a selection criterion. Here a sequence of navigations which lead to an action selection is called an epoch. 

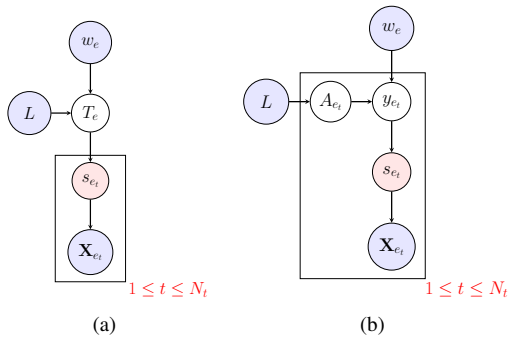
\begin{figure}
\centering
\subfigure[]{
\resizebox{.32\columnwidth}{!}{
        %\begin{centering}
\begin{tikzpicture}[thick,scale=0.9, every node/.style={transform shape}]

\tikzstyle{a} = [rectangle, draw, minimum height=18em, minimum width=10em]
 %\node [response] (Decision) {$s_{e_t}$};
 %\node [response, right = .75cm of Decision] (Decision2) {$D_{k,i+1}$};
%%level1    
 \node [level1] (Position) {\Huge $T_{e}$};
% \node [level1, left = 1cm of Position] (Assignment) {$A_{e_{t}}$};
 \node [response, left = 1cm of Position] (Layout) {\Huge $L$};
 %\node [response, right = .75cm of Position] (Position2) {$y_{e_{t}}$};
 \node [desire, below = 1.5cm of Position] (Decision) {\Huge $s_{e_{t}}$};
 %\node [response, right = .75cm of Decision] (Decision2) {$s_{e_{t}}$};
%%level1    
\node [response, above  = 1.5cm  of  Position] (Context) {\Huge $w_e$};
%\node [level1, above  =1.5cm  of  Position] (State) {$v_e$};
\node [response, below = 1.5cm of  Decision] (EEG) {\Huge $\mathbf{X}_{e_{t}}$};
%\node [level1, below = 1.5cm of  Decision2] (EEG2) {$\mathbf{x}_{e_t}$};
%level1
\draw [arrow] (Context) -- (Position);
\draw [arrow] (Layout) -- (Position);
\draw [arrow] (Position) -- (Decision);
%\draw [arrow] (Layout) -- (Assignment);
%\draw [arrow] (State) -- (Decision2);
%\draw [arrow] (Context) -- (State);
\draw [arrow] (Decision) -- (EEG);
%\draw [arrow] (Decision2) -- (EEG2);
\node[a,fit =(EEG) (Decision) ,label={[label distance=0, text=red]300: \Huge $1\leq t \leq N_t$}] (container) {};

\end{tikzpicture}
}
} 
\subfigure[]{
\resizebox{.42\columnwidth}{!}{
        %\begin{centering}
\begin{tikzpicture}[thick,scale=0.9, every node/.style={transform shape}]

\tikzstyle{a} = [rectangle, draw, minimum height=28em, minimum width=17em]
 %\node [response] (Decision) {$s_{e_t}$};
 %\node [response, right = .75cm of Decision] (Decision2) {$D_{k,i+1}$};
%%level1    
 \node [level1] (Position) {\Huge $y_{e_{t}}$};
 \node [level1, left = 1cm of Position] (Assignment) {\Huge $A_{e_{t}}$};
 \node [response, left = 1cm of Assignment] (Layout) {\Huge $L$};
 %\node [response, right = .75cm of Position] (Position2) {$y_{e_{t}}$};
 \node [desire, below = 1.5cm of Position] (Decision) {\Huge $s_{e_{t}}$};
 %\node [response, right = .75cm of Decision] (Decision2) {$s_{e_{t}}$};
%%level1    
\node [response, above  = 1.5cm  of  Position] (Context) {\Huge $w_e$};
%\node [level1, above  =1.5cm  of  Position] (State) {$v_e$};
\node [response, below = 1.5cm of  Decision] (EEG) {\Huge $\mathbf{X}_{e_{t}}$};
%\node [level1, below = 1.5cm of  Decision2] (EEG2) {$\mathbf{x}_{e_t}$};

%level1
\draw [arrow] (Context) -- (Position);
\draw [arrow] (Assignment) -- (Position);
\draw [arrow] (Position) -- (Decision);
\draw [arrow] (Layout) -- (Assignment);
%\draw [arrow] (State) -- (Decision2);
%\draw [arrow] (Context) -- (State);

\draw [arrow] (Decision) -- (EEG);
%\draw [arrow] (Decision2) -- (EEG2);

\node[a,fit =(EEG) (Assignment) (EEG) (Position),label={[label distance=0, text=red]300: \Huge $1\leq t \leq N_t$}] (container) {};

\end{tikzpicture}
}
}
\caption{Two proposed and compared probabilistic graphical models.(a) joint inference. (b) marginalizing the estimated action probabilities.}%
\label{fig:GraphicalModel}%
\end{figure}

In this section we introduce two probabilistic graphical models used for navigation and inference. In both models, prior probabilities over the nodes
are recursively updated as the user is navigating throughout the graph in each trial. The goal, is to estimate the next command or immediate action, $s_{e_t}$, at epoch $e$ and trial $t$. Here the context information, $\omega_e$, defines a prior distribution over the space of nodes. Moreover, $X_{e_t}$ is the EEG evidence corresponding to $s_{e_t}$, and $L$  represents the layout of graph node. In graphical model demonstrated in Fig.~\ref{fig:GraphicalModel} (a), $T_e$ represents the true state of the system in epoch $e$. Fig.~\ref{fig:GraphicalModel} (b) shows the graphical model introduced in our previous related work~\cite{nezamfar2016flashtype} in which $y_{e_t}$ is the desired pointer location at trial $t$ of epoch $e$; and $A$ represents a particular action assignment on the graph. 
Two different approaches are exploited for estimation of the next command, $s_{e_t}$, in each graphical model. According to our earlier work in~\cite{nezamfar2016flashtype}, we have the following proportion for graphical model (b).
\begin{equation}
    \begin{split}
    \small
    & P(s_{e_t}|\{\mathbf{X}_{e_{i}}\}_{i=1}^{t-1},\mathbf{X}_{e_t},w_e,L)\propto P(\mathbf{X}_{e_t}|s_{e_t}) \sum_{y_{e_t}} P(s_{e_t}|y_{e_t}) \\
    & \ \times \sum_{A_{e_t}} \left[ P(y_{e_t}|A_{e_t},w_e,{\{\mathbf{X}_{e_i}\}}_{i=1}^{t-1},L)P(A_{e_t}|L)\right]
    \end{split}
    \label{eq:posteriorAtt}
    \normalsize
\end{equation}

%In graphical model (a) the joint probability of action decision and true state of the system given the observed EEG evidence and context information is maximized. On the other hand, In graphical model (b), the posterior probability of states given observed random variables is maximized.	Details of equations for graphical model (b) can be found in~\cite{nezamfar2016flashtype}.

%Here we expand equations related to the proposed graphical model shown in Fig.~\ref{fig:graph} (a). 
In graphical model (a) the joint probability of action decision and true state of the system given the observed EEG evidence and context information is maximized. Assume that $\{\mathbf{X}_{e_{i}}\}_{i=1}^{t},\mathbf{X}_{e_t},w_e$ and $L$ are observed. Therefore, the joint probability of action decision and true state of the system given all observed variables can be written as, 
%Then according to the assumptions presented in the probabilistic graphical model (a) of the system we have,
\begin{equation}
\begin{split}
\small
 & \ P(s_{e_t},T_e|{\{X_{e_i}\}}_{i=1}^{t-1},X_{e_t},w_e,L) \\
 & \ \propto P(s_{e_t},T_e,X_{e_t}|{\{X_{e_i}\}}_{i=1}^{t-1},w_e,L) \\
 & \ =  P(X_{e_t}|s_{e_t}) \times P(s_{e_t},T_e|{\{X_{e_i}\}}_{i=1}^{t-1},w_e,L) \\
 & \ =  P(X_{e_t}|s_{e_t}) \times P(s_{e_t}|T_e) \\
 & \ \times P(T_e|{\{X_{e_i}\}}_{i=1}^{t-1},w_e,L)
\end{split}
\label{equation:1}
\normalsize
\end{equation}
% Expanding probability of the true state of the graph in each epoch will result the following form,
In equation \eqref{equation:1}, $P(T_e|{\{X_{e_i}\}}_{i=1}^{t-1},w_e,L)$ represents the probability over the state space estimated from the context and observed EEG in earlier trials. This probability mass function can be updated after each trial as shown in \eqref{eq3}.
%Then assuming, $P(T_e|{\{X_{e_i}\}}_{i=1}^{t-1},w_e,L)$, as the probability of true state of the graph according to the context information and observed EEG evidence can be written as follows, 
 \begin{equation}
\begin{split}
\small
 & \ P(T_e|{\{X_{e_i}\}}_{i=1}^{t-1},w_e,L) \\
 & \ \propto P(T_e|w_e,L) \times \sum_{s_{e_{t-1}}} P(X_{e_{t-1}}|{s_{e_{t-1}}}) P({s_{e_{t-1}}}|T_e) \times \cdots \\
 & \ \times \sum_{s_{e_1}} P(X_{e_1}|{s_{e_1}}) P({s_{e_1}}|T_e)\\
 & \ = P(T_e|{\{X_{e_i}\}}_{i=1}^{t-2},w_e,L) \sum_{s_{e_{t-1}}} P(X_{e_{t-1}}|{s_{e_{t-1}}}) P({s_{e_{t-1}}}|T_e)
\label{eq3}
\end{split}
\normalsize
\end{equation}
For more detail on derivations for Equation~\ref{eq3} please see Appendix~\ref{app:deriv}. In the \eqref{eq3},
\begin{itemize}
\item $P(\mathbf{X}_{e_t}|s_{e_t})$, represents likelihood of EEG evidence given each possible class of actions in trial $t$ of epoch $e$. 
%represents the EEG observation likelihood for a particular command.
\item $P(s_{e_t}|T_e)$, is the probability of taking each action class given true state of the graph in trial $t$ of epoch $e$.
%by which the user selects a particular command to take the specific character from set of $T_e$. 
\end{itemize}
\subsection{Target \emph{Selection} Criteria}
\label{sec:tsc}
Two decision criteria for epoch conclusion was utilized; first the user need to choose a “Select” command, second, if the ratio of the current pointer location probability, to the next most probable location exceeds a predefined threshold ($\frac{P(T_e=currentLocation|{\{X_{e_i}\}}_{i=1}^{t-1},w_e,L)}{P(T_e=SecondMostProbableLoc|{\{X_{e_i}\}}_{i=1}^{t-1},w_e,L)}>Tr$) the system selects current node. In this manuscript we refer to this condition as Probabilistic \emph{Selection} Criterion (PSC). 

\section{Simulation Results}
\subsection{Simulation Study Design}
\label{sec:simul}
In this study, a code visually evoked potential (c-VEP) based BCI gridded keyboard, FlashType,~\cite{nezamfar2016flashtype} is utilized to assess the effectiveness of; (1) graphical models described in Section~\ref{sec:pgms} and (2) target \emph{Selection} criteria explained in Section~\ref{sec:tsc}. Here each character on the keyboard is represented by a node of degree four (m=4) on a graph with 35 nodes. A 6-gram language model provides prior context information for each character\footnote{Note that the word suggestion feature of the FlashType~\cite{nezamfar2016flashtype} was turned off during this simulation.}. Typing each character consist of a sequence of cursor movements or actions in an epoch finalized by selecting a target character. Four checkerboards located at the corners of the gridded keyboard, are used as stimuli.
Each checkerboard is mapped to a different action moving the cursor on the gridded keyboard.
%Subjects gaze at each checkerboard depending on their desired move direction and EEG recordings are used as brain evidence to estimate likelihood of the intended direction of user, $P(\mathbf{X}_{e_t}|s_{e_t})$.

% To compare the performance of the system using each approach, a 
A Monte Carlo based simulation is performed over typing ten words from individual sentences, called a session, each with 30 runs.  The words come from five different difficulty levels based on language model prediction. Levels with higher difficulty imply lower prediction probability based on language model and vice versa. 
%\REVO{We want to evaluate the effect of the prior context information's confidence (language model reliability) on typing.}
Typing the following ten words "the, and, with, will, seat, between, seen, please, buys, makeup" is defined at a simulation run. The level of difficulty increases by one after every two words.

Seven pre-recorded c-VEP calibration data sets with high, average, and low classification accuracies have been used to produce realistic EEG features for simulation~\cite{nezamfar2016flashlife,orhan2016probabilistic}. The accuracy of each data set has been estimated by employing a leave-one-out cross validation among different selection options.

To simulate typing each character, the system assumes subject's intended navigation sequence in each epoch is equal to the path with minimum number of cursor movements toward the desired character. Therefore, in each trial true action command is chosen based on the determined minimum path to the desired character. Next, samples from true action command class are drown based on estimated class conditional distributions. The set of likelihood scores for i.i.d drawn samples are then calculated using the class conditional densities.

Three probabilistic models for estimating subject's next intended direction or command are evaluated and compared. In graphical model (a) joint probability of true state and actions $P(s_{e_t},T_e|{\{X_{e_i}\}}_{i=1}^{t-1},X_{e_t},w_e,L)$, in graphical model (b) posterior probability of states, $P(s_{e_t}|{\{X_{e_i}\}}_{i=1}^{t-1},X_{e_t},w_e,L)$, and in the case of using no graphical model the likelihood probability of states, $P(\mathbf{X}_{e_t}|s_{e_t})$, are calculated. The calculated probabilities in each case need to reach a certain \emph{Confidence Threshold}, for the corresponding command to be chosen. Otherwise, another set of likelihood scores is randomly drawn from the conditional densities and conditional probabilities for each model are calculated again. This procedure continues until the \emph{Confidence Threshold} or maximum allowed number of repetitions (5 in this simulation) has been reached.

\subsection{Model Comparison Results}
Two PGMs (probabilistic graphical models) explained in Section~\ref{sec:pgms} and two target \emph{Selection} criteria explained in Section~\ref{sec:tsc} are compared in this section. A PGM along with a target \emph{Selection} criterion can be used for each typing approach. Different pairs of PGMs and \emph{Selection} criterion are used for comparision.
%Following pairs of PGMs and a target selection criterion in addition to the case of using no PGM, form all the available approaches to be compared: PGM (a) with "Select" command, PGM (b) with "Select" command, PGM (a) with PSC, PGM (b) with PSC, and No PGM used.

\begin{figure}
\centering
\includegraphics[width=.8\columnwidth]{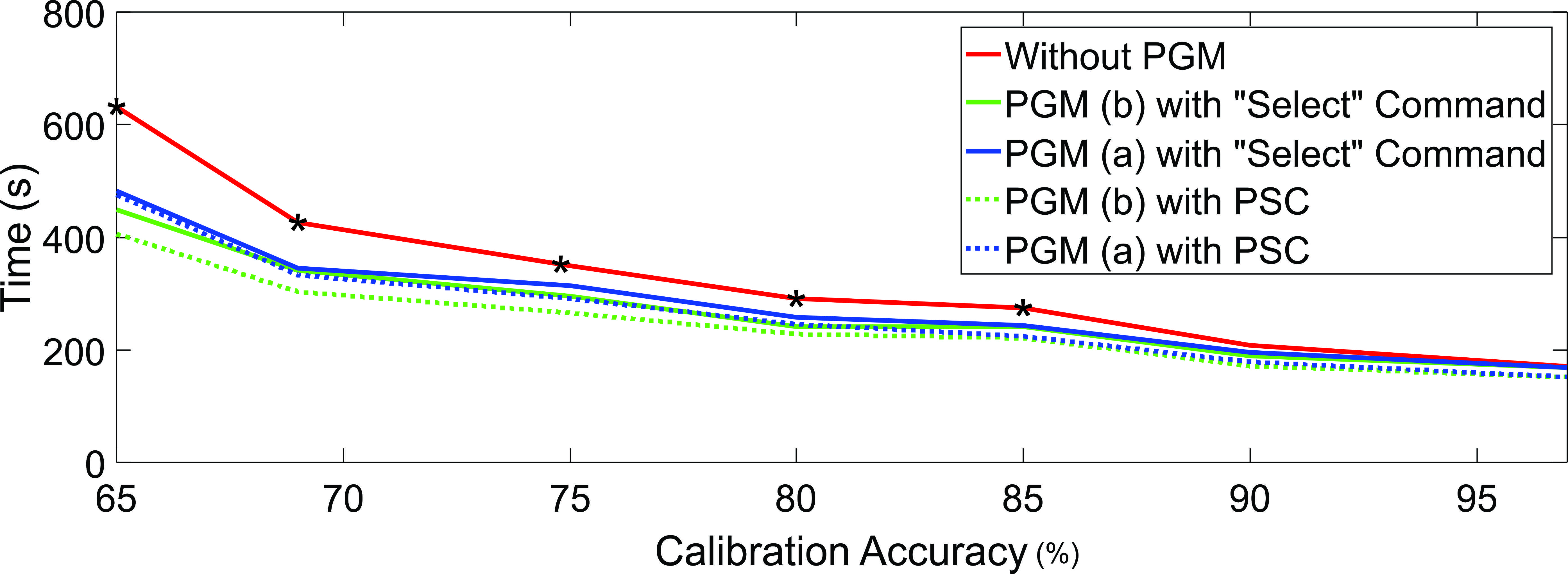}
\caption{Average estimated time based on 30 Monte Carlo simulations, to type ten words, employing different PGMs. * shows the significant decrease when models are employed compare to Without PGM.}
\label{fig:time}
\end{figure}

Using the simulation study explained in section~\ref{sec:simul}, effect of employing each approach on average typing duration of each session is illustrated in Fig.~\ref{fig:time}. For each pre-recorded calibration data sets, total time for completing each session (typing ten words) was first calculated and then averaged over all 30 simulated runs. Note that each likelihood score in simulation, will require 1.05 second of EEG observation in a real experiment scenario. This figure indicates using PGMs enhances the typing speed specifically for the users with low calibration  performance. Time needed to complete a session was significantly decrease when either graphical model (a) or (b) was used compared to not using a graphical model, the significant difference is calculated based on a t-test with   $\alpha$ threshold of 0.001 ($p<0.001$). The significant decrease is observed in data sets with calibration accuracies less than 90\%.
Overall, adding the PGM (b) along with PSC provides the highest typing speed improvement.

\begin{figure}
\centering
\includegraphics[width=.8\columnwidth]{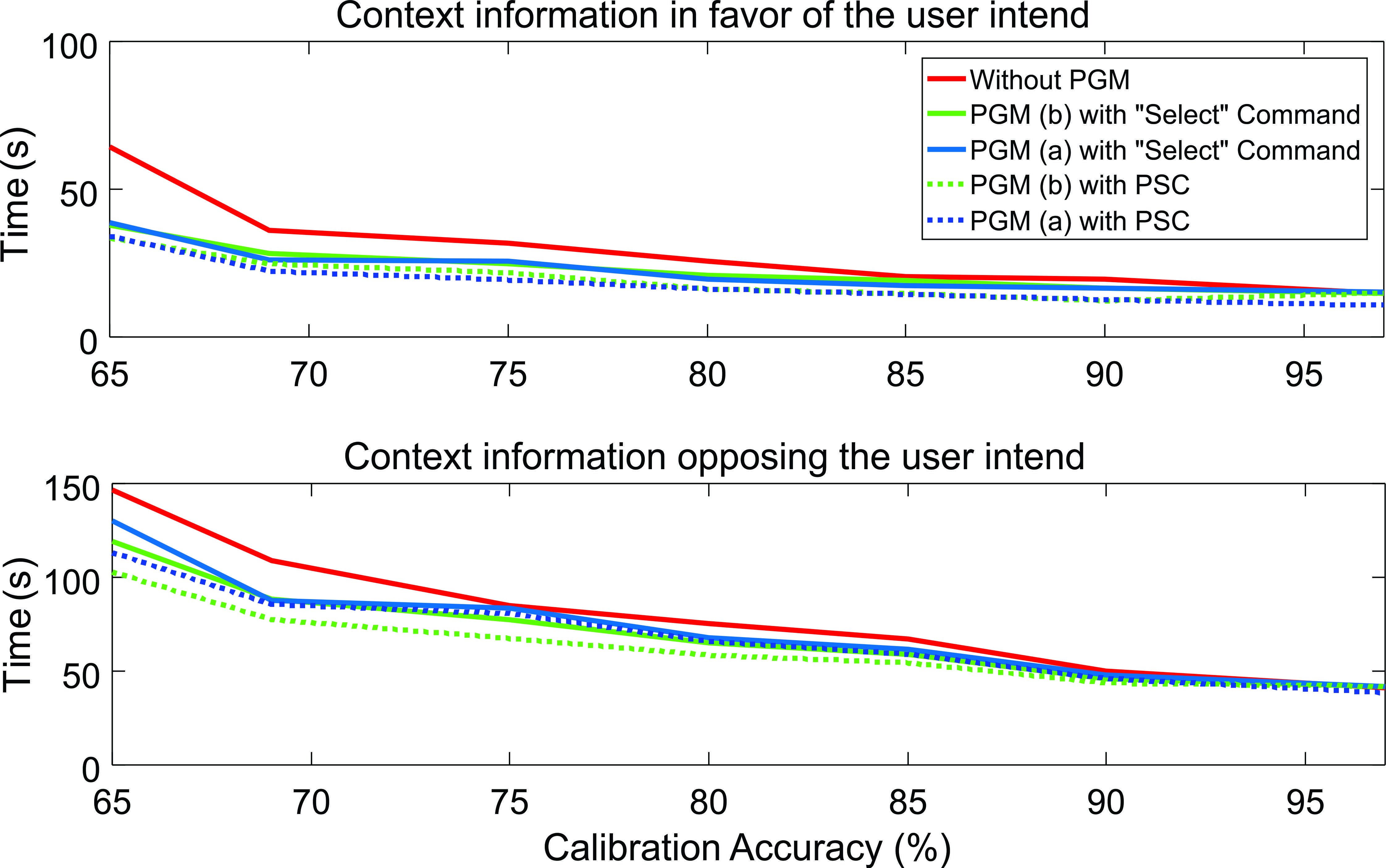}
\caption{Average estimated time based on 30 Monte Carlo simulations, to type two words. Upper plot: Context information in favor of the user intend. Lower: Context information opposing of the user intend.}
\label{fig:Context}
\end{figure}

The difference between graphical model (a) and (b) can be seen in equation~\ref{eq:posteriorAtt} and \ref{equation:1}. While  graphical model (b) is marginalizing out the desired pointer location over nodes, graphical model (a) chooses the maximum joint probability of the true state of nodes and immediate actions. Hence, the context information has a stronger effect on graphical model (a). Figure~\ref{fig:Context} emphasises this difference by showing the time needed to type two complete words, when context information is in favor of user intend (upper plot) and opposing the user intend (lower plot). When  context is in favor of user intend graphical model (a) performs better. On the other hand, when the context is not in favor of user intend graphical model (b) performs better.
% \vspace{-2mm}
\section{Discussion}
% \vspace{-2mm}
In this study, we introduced two PGMs which use context information and previously observed EEGs in addition to the EEG recorded during the current trial. Our simulation results show PGMs along with PSC can significantly enhance typing speed especially for users with poor EEG classification performance.
%In the BCI based graph traversal, user navigates the cursor to the desired option in the gridded space of options. Each option has been given a specific prior probability based on context information. We showed that using the introduced graphical models can decrease decision time and make decisions more accurate. 

\begin {figure}
\centering
\includegraphics[width=0.8\columnwidth]{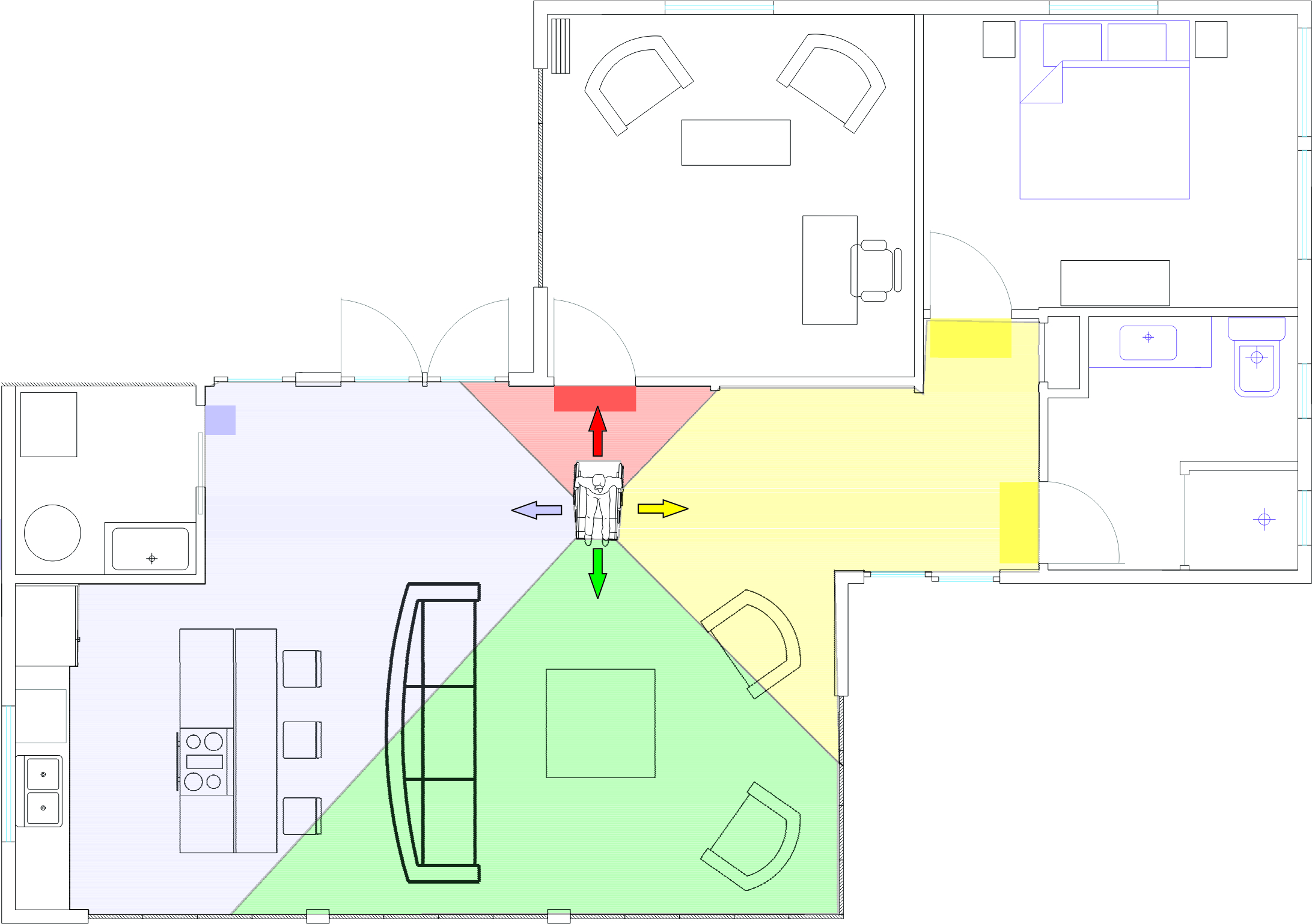}
\caption{The home plan example and continues presence probability. The presence probability and previous decisions could be fused using the introduced graphical models.}
\label{fig:Home Plan}
%\vspace{-2mm}
\end{figure}

Similarly, these models can be employed in SSVEP based wheelchair navigation. Fig.~\ref{fig:Home Plan} shows the example plan of a home environment as a set of user presence. Each location on this plan has a specific user presence probability. Assuming four SSVEP options (turning left, turning right, forward, and backward) to move the wheelchair, context information for each of these discrete options is extracted based on the spatial integration of possible targets in each direction. 

%Second, in wheelchair navigation, similar to graph traversal, previous decisions are affecting the next decision. Using introduced graphical models to fuse EEG information, previous decisions, and presences probability (as the context information, $\omega_e$) will increase the reliability of decisions in BCI wheelchair controlling. 

%\vspace{-2mm}
%\section{CONCLUSIONS}
% \vspace{-2mm}
%In this study, we introduced two PGMs which use context information and previously observed EEGs in addition to the EEG recorded during the current trial. Our simulation results show PGMs along with PSC significantly enhance the typing speed especially for users with poor EEG classification performance.
\section*{APPENDIX}
According to Fig.~\ref{fig:GraphicalModel}.a the probability of the true state given observed random variables can be written as:
\label{app:deriv}
\begin{equation}
\small
\begin{split}
 & \ P(T_e|{\{X_{e_i}\}}_{i=1}^{t-1},w_e,L) \\
 & \ =  \sum_{\{s_{e_i}\}_{i=1}^{t-1}} P(T_e,s_{e_i}|{\{X_{e_i}\}}_{i=1}^{t-1},w_e,L) \\
 & \ \propto  \sum_{\{s_{e_i}\}_{i=1}^{t-1}} P(T_e,s_{e_i},{\{X_{e_i}\}}_{i=1}^{t-1}|w_e,L) \\ 
 & \ =  \sum_{\{s_{e_i}\}_{i=1}^{t-1}} P(T_e,s_{e_i},{\{X_{e_i}\}}_{i=1}^{t-1}|T_e) \times P(T_e|w_e,L) \\
 & \ =  P(T_e|w_e,L) \times \sum_{\{s_{e_i}\}_{i=1}^{t-1}} [P({\{X_{e_t}\}}_{i=1}^{t-1}|s_{e_i}) \times P(s_{e_i}|T_e)] \\
 & \ = P(T_e|w_e,L) \times \sum_{\{s_{e_i}\}_{i=1}^{t-1}} \prod_{i=1}^{t-1} P(X_{e_i}|{s_{e_i}}) P({s_{e_i}}|T_e) 
%  & \ = P(T_e|w_e,L) \times \sum_{s_{e_{t-1}}} P(X_{e_{t-1}}|{s_{e_{t-1}}}) P({s_{e_{t-1}}}|T_e) \times \cdots \\
%  & \ \times \sum_{s_{e_1}} P(X_{e_1}|{s_{e_1}}) P({s_{e_1}}|T_e)
\label{equation:2}
\end{split}
\normalsize
\end{equation}
Equation~\ref{equation:2} can be evaluated in recursive mode.
\begin{equation}
\small
\begin{split}
       & \ P(T_e|{\{X_{e_i}\}}_{i=1}^{t-1},w_e,L) \\
       & \propto P(T_e|{\{X_{e_i}\}}_{i=1}^{t-2},w_e,L) \times \sum_{s_{e_{t-1}}} P(X_{e_{t-1}}|{s_{e_{t-1}}}) P({s_{e_{t-1}}}|T_e)
\label{equation:3}
\end{split}
\normalsize
\end{equation}
Which is equal to equation~\ref{eq3}.

% \begin{thebibliography}{99}

% \bibitem{c1} G. O. Young, �Synthetic structure of industrial plastics (Book style with paper title and editor),� 	in Plastics, 2nd ed. vol. 3, J. Peters, Ed.  New York: McGraw-Hill, 1964, pp. 15�64.

% \end{thebibliography}

\bibliographystyle{IEEEtran}
\bibliography{refs}

\end{document}